\documentclass[lettersize,journal]{IEEEtran}
\usepackage{amsmath,amsfonts}
\usepackage{algorithmic}
\usepackage{algorithm}
\usepackage{array}
\usepackage[caption=false,font=normalsize,labelfont=sf,textfont=sf]{subfig}
\usepackage{textcomp}
\usepackage{stfloats}
\usepackage{url}
\usepackage{verbatim}
\usepackage{graphicx}
\usepackage{cite}
\usepackage{bm}
\usepackage{xcolor}
\usepackage{booktabs}

\begin{document}

\title{3DLS: A 3D Logic-Stacked Architecture for Disaggregated LLM Serving}

\author{Jaehun Lee, In-Jun Jung, and Joo-Young Kim, \IEEEmembership{Senior Member, IEEE}

\thanks{\copyright~2026 IEEE. This is the author's accepted manuscript of an article accepted for publication in \textit{IEEE Computer Architecture Letters}. DOI: 10.1109/LCA.2026.3709108. Personal use of this material is permitted. Permission from IEEE must be obtained for all other uses.}
\thanks{Jaehun Lee is with the Graduate School of System Architect, KAIST, Daejeon 34141, South Korea (e-mail: jaehunlee@kaist.ac.kr). }

\thanks{In-Jun Jung and Joo-Young Kim are with the School of Electrical Engineering, KAIST, Daejeon 34141, South Korea (e-mail: injun@kaist.ac.kr; jooyoung1203@kaist.ac.kr).}
}



\maketitle
\begin{abstract}
Large language model (LLM) serving increasingly combines prefill-decode (PD) disaggregation with tensor parallelism (TP) to support large models and long contexts. In conventional 2D/2.5D chiplet architectures, layer-wise prefill-to-decode KV-cache transfer decode-side TP collectives share the same lateral die-to-die (D2D) interconnect, creating mixed-traffic contention on the decode critical path. This contention increases communication latency, prolongs token generation intervals, and degrades end-to-end serving performance. We propose 3DLS, a logic-on-logic 3D-stacked chiplet architecture that separates traffic classes by routing KV-cache transfers through vertical interconnects while preserving decode-side TP collectives on the lateral D2D fabric. 3DLS achieves up to 1.49$\times$ throughput and 60.2\% lower end-to-end (E2E) latency over the shared-fabric planar baseline, and still achieves up to 1.17$\times$ throughput and 31.4\% lower E2E latency over a workload-aware priority-managed planar baseline. These results highlight that physical isolation is an important design principle for future chiplet-based PD-disaggregated LLM serving systems.
\end{abstract}


\section{Introduction}

\IEEEPARstart{L}{arge} language model (LLM) serving is shifting toward larger models and longer context windows, increasing compute and memory resource demands. At the same time, satisfying these demands with monolithic large-die accelerators has become increasingly difficult due to the rising cost of advanced process nodes, yield challenges, and reticle-size limits. These trends make chiplet-based integration a promising solution for scaling LLM inference hardware beyond the practical limits of monolithic designs~\cite{reticlelimit}, making on-package communication a major concern for serving performance.

Modern LLM inference consists of two phases with distinct characteristics: a compute-intensive prefill phase and a memory-bound decode phase. Because these phases differ in resource demands, recent serving systems increasingly adopt prefill-decode (PD) disaggregation~\cite{distserve, pdserve}, assigning each phase to a dedicated resource pool to improve efficiency and utilization. Meanwhile, serving large models continues to rely on tensor parallelism (TP)~\cite{megatron} to meet memory-capacity and throughput requirements. However, TP introduces frequent collective operations whose overhead grows with model scale and TP degree. As a result, a TP-enabled PD-disaggregated serving system must simultaneously support two heterogeneous traffic classes: KV-cache transfer from prefill to decode, and latency-critical decode-side collective communication.

\begin{figure}
    \centering
    \includegraphics[width=0.44\textwidth]{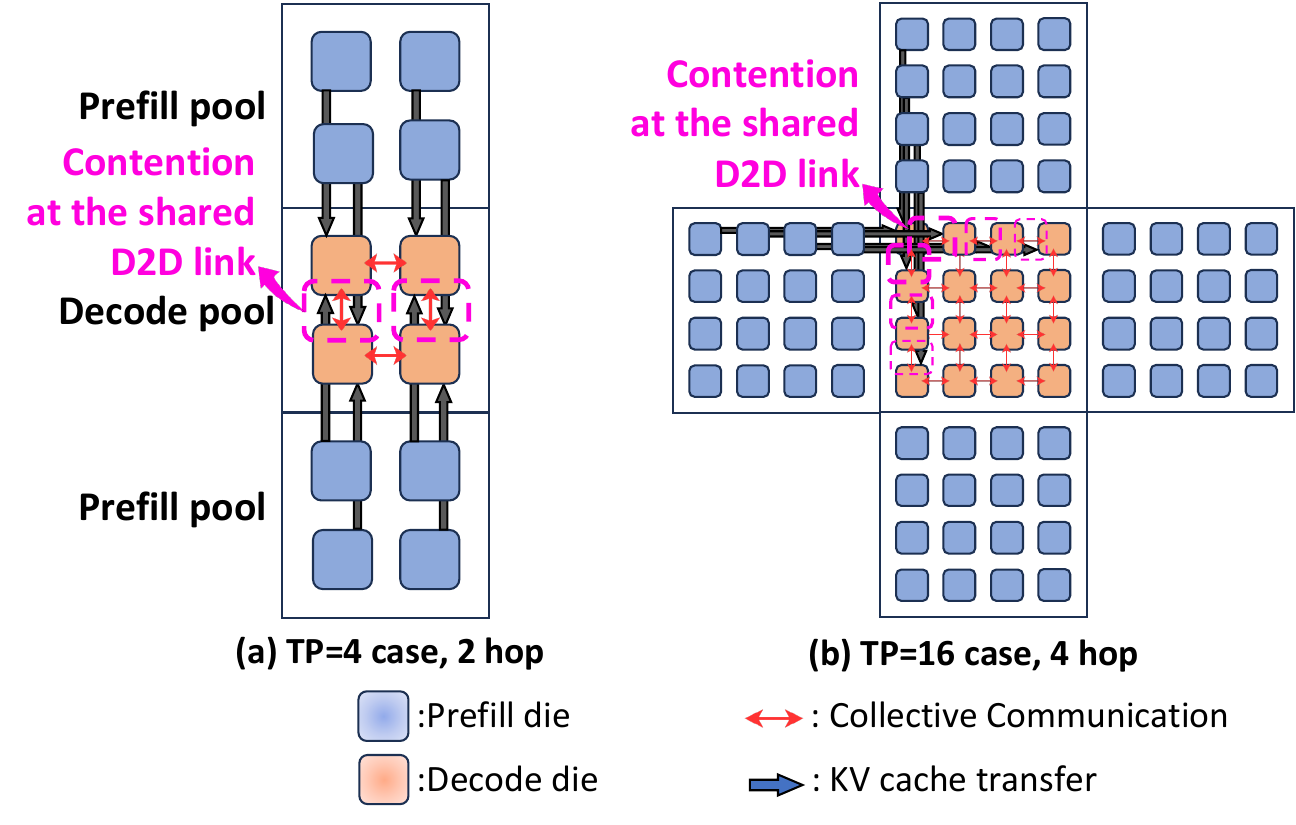}
    \caption{Traffic contention in conventional 2D/2.5D chiplet-based PD-disaggregated serving. (a) TP=4 case with 2-hop KV cache transfer; (b) TP=16 case with 4-hop KV cache transfer.} 
    \label{fig:fig1}
\end{figure}

{Fig.~1 shows that when PD-disaggregated serving with TP} is mapped onto a conventional 2D/2.5D chiplet architecture~\cite{wscllm}, KV-cache transfer and decode-side TP collectives are forced to share the same lateral die-to-die (D2D) interconnect. The asymmetric pool ratios follow decode-centric placement for large-batch, long-context serving~\cite{wscllm}: Fig.~1(a) assumes a 2:1 prefill:decode pool ratio, while Fig.~1(b) uses 4:1 for higher TP to align layer-wise KV arrivals across shards. Simply increasing the number of decode dies does not resolve this contention: distributing TP shards across more dies introduces arrival-time skew that delays decode, while KV-cache transfer and decode-side AllReduce still share the same lateral D2D fabric. However, prior work has not directly addressed this contention, instead focusing on design-space exploration of network topology, scheduling, placement, and resource allocation~\cite{pdconstraint, wscllm}. While such optimizations can indirectly mitigate contention, they do not eliminate its root cause: the lack of physical isolation between heterogeneous traffic classes.

\IEEEpubidadjcol
To address this limitation, we propose 3DLS, a logic-on-logic 3D-stacked chiplet architecture for PD-disaggregated LLM serving with TP. 3DLS places multiple prefill pools on a top tier and multiple decode pools, together with the decode-side D2D fabric, on a bottom tier. Prefill-to-decode KV-cache transfer is routed through dedicated vertical 3D interconnects (e.g., TSVs or hybrid bonding), while decode-side TP collectives remain on the lateral D2D interconnect. The key idea is not to use 3D integration merely for higher bandwidth, but to co-design the stacked architecture and communication paths so that the two heterogeneous traffic classes are physically isolated. In summary, this paper makes three contributions. First, it characterizes the mixed-traffic contention as a key bottleneck in the chiplet-based LLM serving system. Second, it proposes 3DLS, which separates the KV-cache transfer path from the decode-side collective path. Third, it demonstrates that physical traffic isolation reduces interconnect contention, improving both throughput and end-to-end latency.

\begin{figure}
    \centering
    \includegraphics[width=0.49\textwidth]{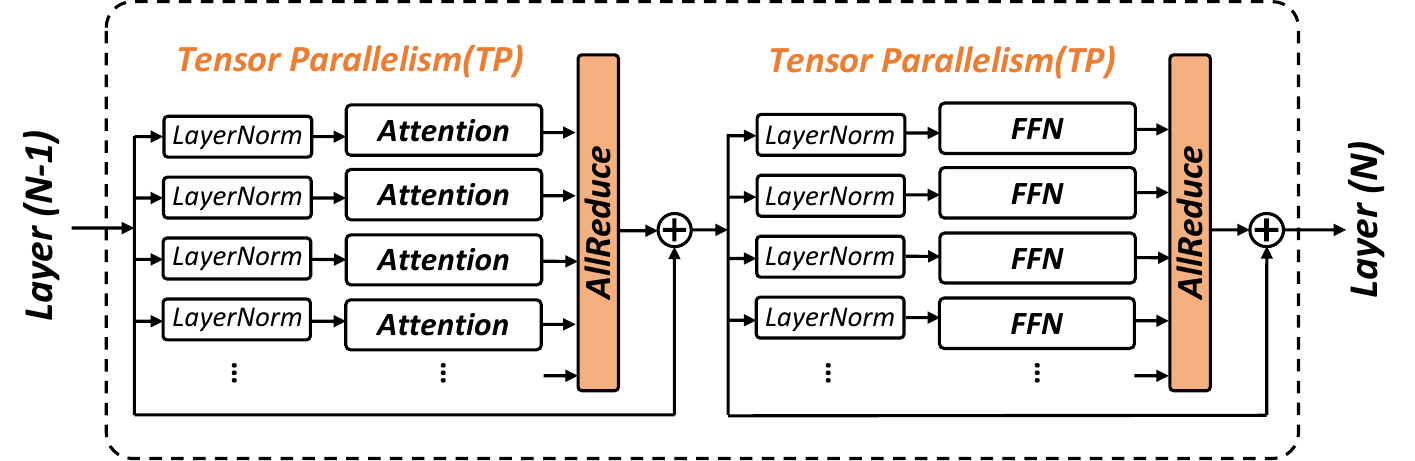}
    \caption{Communication operations in a representative tensor-parallel transformer decoder layer. Decode-side tensor parallelism introduces two AllReduce operations on the layer critical path.} 
    \label{fig:fig2}
\end{figure}

\section{Motivation and Problem Characterization}


We refer to the conventional 2D/2.5D chiplet architecture in Fig.~1 as Naive Planar. In this baseline, KV-cache transfer (KVT) and decode-side TP collectives share the same lateral D2D interconnect. This contention prolongs decode latency because every AllReduce (AR) can be delayed by concurrent KV-cache traffic. We adopt layer-wise transfer~\cite{splitwise} as the baseline, since it pipelines KVT with prefill computation and represents the more latency-optimized transfer strategy.
{In a tensor-parallel transformer decode layer, as shown in Fig.~2, TP places two AR operations on the forward critical path: one after self-attention and one after the FFN block~\cite{megatron}. Because these collectives occur at every layer for every generated token, any D2D delay is repeatedly exposed in decode time.
Fig.~3 shows that each completed prefill layer emits a KV block to the decode pool while an in-flight decode request performs layer-wise AR on the same lateral fabric. Thus, KVT can queue ahead of or alongside AR, stretching the token step.

\begin{figure}
    \centering
    \includegraphics[width=0.42\textwidth]{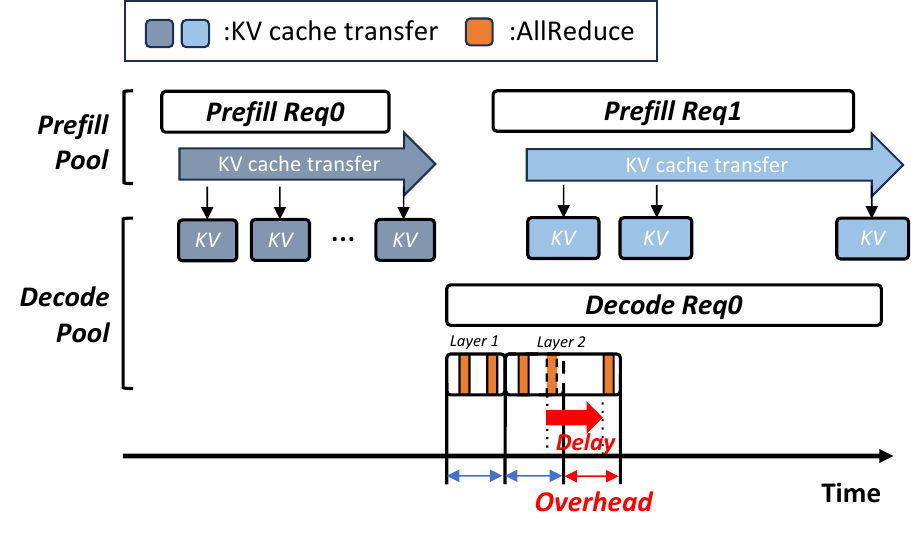}
    \caption{Timing diagram of conflict between layer-wise KV-cache transfer and decode-side AllReduce in 2D/2.5D chiplet-based PD-disaggregated serving. The conflict delays AllReduce and the token generation time.} 
    \label{fig:fig3}
\end{figure}

\begin{figure}
    \centering
    \includegraphics[width=0.49\textwidth]{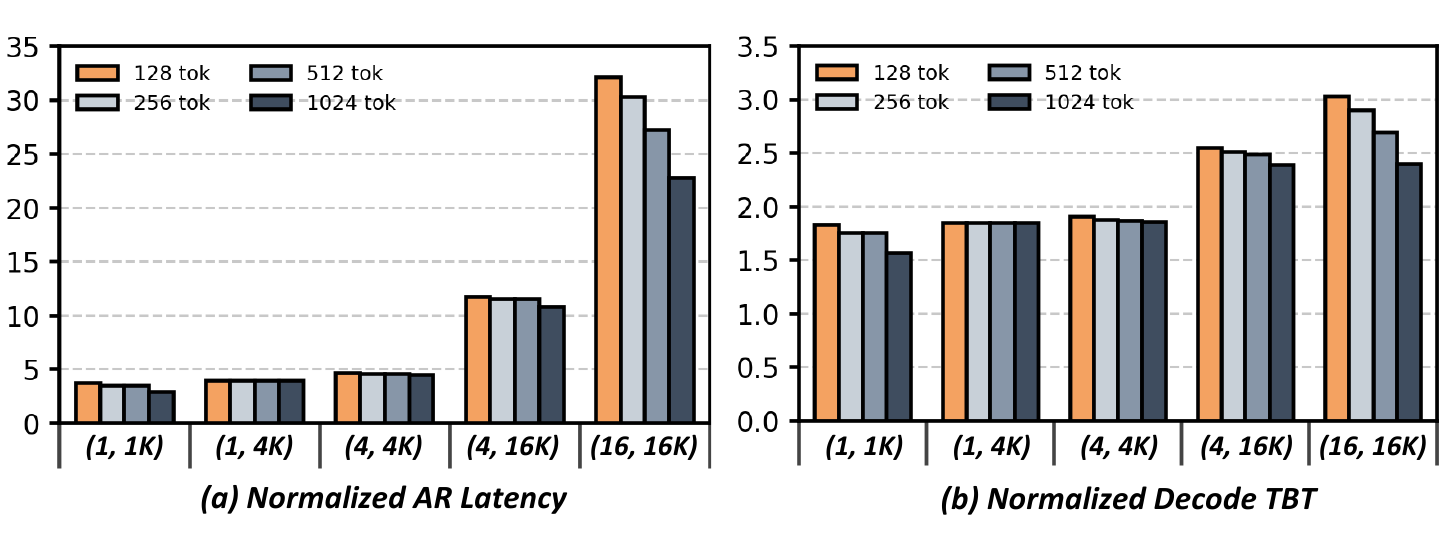}
    \caption{Quantitative impact of KVT/AR contention in Naive Planar. (a) AR latency increases with contention compared with the contention-free case. (b) Longer decode TBT due to the increase of the AR latency.} 
    \label{fig:fig4}
\end{figure}

Fig.~4 quantifies KVT/AR contention in the Naive Planar setting. The experiment uses OPT-175B~\cite{opt}, TP=16, and 512 GB/s D2D bandwidth. We define KV load as (batch size, context) and evaluate five regimes: (1, 1K), (1, 4K), (4, 4K), (4, 16K), and (16, 16K). We choose 128 output tokens as a controlled characterization point because it is close to the median generated length of the Azure conversation trace (129 tokens)~\cite{azure}. We additionally sweep output token length from 128 to 1024 tokens to evaluate output-length sensitivity. The results are normalized to a contention-free baseline under the same model, KV-load, output token length, and D2D bandwidth setting. AllReduce latency denotes the request-level accumulated AllReduce latency during decode, while decode time-between-token (TBT) denotes the average per-token decode latency. At the highest-KV-load point (batch size=16, input length=16K), accumulated AR latency increases by 32.1$\times$ from 510ms to 16.37s, and average decode TBT increases by 3.03$\times$ from 61.11ms to 184.92ms at 128 output tokens, and still shows 22.8$\times$ AR latency and 2.39$\times$ decode TBT at 1024 output tokens. Thus, longer outputs reduce the relative severity of KVT/AR contention, but do not remove the critical-path penalty.

These results show that the bottleneck is shared-path contention, not simply insufficient average bandwidth. Priority-management mechanisms, such as quality-of-service (QoS)-aware scheduling and virtual channels (VC), can mitigate this contention and improve over a naive shared fabric. However, because KVT and AR still consume the same physical links, priority management reallocates interference rather than eliminating the shared-path contention: prioritizing AR can delay KV handoff, while prioritizing KVT can delay AR. D2D bandwidth over-provisioning can also reduce contention, but it is not a free design knob in 2.5D systems because D2D bandwidth is constrained by die-edge/interposer shoreline resources and competes with other package I/O resources, including HBM capacity and aggregate memory bandwidth, which are critical for memory-bound LLM inference. 3DLS addresses this root cause through physical path isolation, mapping KVT and AR onto disjoint paths.
\section{Proposed Architecture: 3DLS}

\begin{figure}
    \centering
    \includegraphics[width=0.49\textwidth]{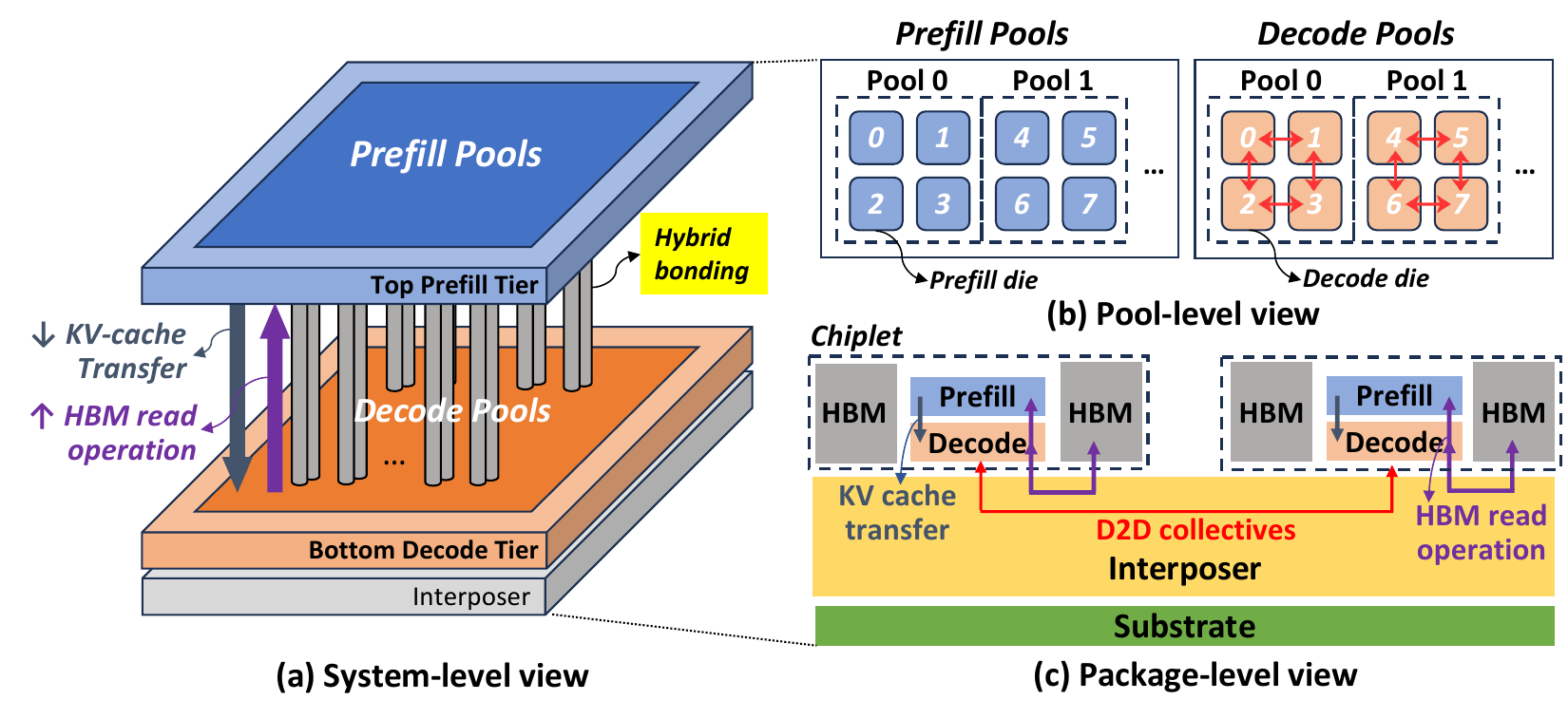}
    \caption{3DLS Overall Architecture. (a) System-level view; (b) Pool-level view, where asymmetric prefill:decode ratios denote logical pool allocation (c) Package-level view, showing top-to-bottom KVT, bottom-to-top HBM reads, and lateral decode-side D2D collectives.}
    \label{fig:fig5}
\end{figure}

3DLS is a logic-on-logic 3D-stacked chiplet architecture that physically separates prefill-to-decode KV-cache transfer from decode-side collective communication. At the system level, Fig.~5(a), prefill pools reside on the top tier and decode pools reside on the bottom tier. Requests are processed by a prefill pool, and their layer-wise KV blocks are delivered through vertical links to the corresponding decode-side KV-cache ingress domain, while decode-side collectives remain on the bottom-tier lateral 2.5D D2D fabric, e.g., a UCIe-like fabric~\cite{ucie}. This separation removes the shared KVT/AR physical path.
Fig.~5(b) provides a pool-level view. A pool is a logical group of TP-sharded chiplets, and the prefill:decode ratio is a logical pool-allocation ratio rather than a physical die-stacking ratio. Thus, asymmetric ratios such as 2:1 do not imply one-to-one die pairing, stacking multiple prefill dies onto one decode die, or all-to-one KVT into a single decode die; KVT instead follows shard-aligned transfers to matching decode-side KVT ingress endpoints. Higher TP degree or more asymmetric prefill:decode pool ratio increases lateral KVT span and shared-fabric contention in Naive Planar, making the physical separation in 3DLS more beneficial.

At the package level, Fig.~5(c), HBM remains attached to the interposer/bottom-tier side and is shared by the prefill and decode dies. Top-tier prefill dies stream weight tiles, and dataflow-dependent activation tiles if needed, from bottom-tier HBM through vertical links. Following UCIe-3D bidirectional TX/RX bundle organization~\cite{ucie}, bottom-to-top bandwidth serves HBM reads, while top-to-bottom bandwidth carries generated KV blocks. This adds a vertical memory hop, but top-tier HBM reads do not consume the bottom-tier lateral D2D fabric used by decode-side collectives and therefore do not reintroduce KVT/AR contention. Because prefill is compute-heavy and has high arithmetic intensity, these tiled memory streams can be pipelined and overlapped with computation, reducing the impact of the added vertical-hop latency.
3DLS uses UCIe-3D-class vertical interconnect, which provide substantially higher bandwidth density and lower energy-per-bit targets than UCIe-A-class 2.5D D2D links~\cite{ucie}. This allows 3DLS to isolate KVT from decode-side AR without consuming bottom-tier shoreline. By reducing pressure on the limited die-edge/interposer resources that memory interfaces and lateral D2D links compete for in planar architecture~\cite{wscllm}, 3DLS opens a package-level DSE opportunity to balance lateral D2D bandwidth, and HBM capacity/bandwidth, while accounting for phase-dependent lateral/vertical utilization.
However, a practical 3DLS implementation must account for stacking overheads. These overheads include vertical links, bonding interfaces, PHY/control logic, KV-cache ingress resources, and power-delivery structures, which introduce area and power costs. Thus, 3DLS should be designed with thermal-budget-aware placement and power/cooling co-design; Section~IV provides a first-order thermal-envelope check for the evaluated two-tier stack. Package yield is another implementation constraint that can be addressed through known-good-die testing before bonding, bonding-yield screening, and repair/redundancy support for vertical links.

\section{Evaluation}

\subsection{Evaluation Methodology}

We evaluate 3DLS against two planar baselines using a trace-driven in-house simulator that models TP, layer-wise compute and communication, and prefill-to-decode KVT. Naive Planar denotes the conventional 2D/2.5D shared-fabric organization in Fig.~1, where layer-wise KVT and decode-side AR share the same lateral D2D fabric. PM-Planar strengthens this baseline with logical isolation: KVT and AR use separate virtual channels and static weighted bandwidth reservation, but still share the same physical lateral fabric.
Following~\cite{wscllm}, we configure compute capacity by scaling the Tesla Dojo~\cite{dojo} D1 die to a 16$\times$16 configuration and adopt a decode-centric placement with a prefill:decode pool ratio of 2:1. As a first-order check against thermal budget, the evaluated two-tier 3DLS is designed under a reported 200 W/cm$^2$ advanced-cooling envelope for 3D accelerators~\cite{stratum}. For a fair comparison, we adopt an iso-bandwidth configuration: the lateral links in both planar baselines and the 3DLS lateral and vertical links are modeled with the same bandwidth, so that any performance gain reflects traffic-path isolation rather than additional bandwidth from 3D integration. We model the vertical interconnect as full-duplex, 512 GB/s total bidirectional bandwidth, provisioned as 256 GB/s per direction. KVT uses the top-to-bottom direction, while top-tier HBM reads use the bottom-to-top direction. Using the peak prefill throughput of 261.12 TFLOPS, the evaluated upper-bound weight-streaming demand fits within the 256 GB/s bottom-to-top budget.

We evaluate LLaMA3-8B, LLaMA3-70B~\cite{llama}, and OPT-175B~\cite{opt} on Azure conversation (Conv) and code Code) traces~\cite{azure}. Conv is decode-dominant due to longer generations, while Code is prompt/KV-dominant due to longer prompts and shorter generations. To meet memory-capacity needs, we use TP=4, 8, and 16 for the three models, respectively.
For each model and workload, we sweep the load by scaling trace arrival rate, together with prefill batch (PB) and decode batch (DB), and select a representative Naive Planar operating point with a balanced latency-throughput trade-off, and apply the same operating point to PM-Planar and 3DLS. For PM-Planar, we sweep static KVT:AR reservations of 25:75, 50:50, and 75:25 and report the  best workload-level partition:25:75 for Conv and 75:25 for Code. This gives PM-Planar a favorable comparison point rather than forcing one fixed partition across mixed online serving.

\begin{table}[t]
\centering
\refstepcounter{table}
\label{tab:sys_param}
\small
\begin{tabular}{@{}ll@{}}
\multicolumn{2}{c}{\textbf{Table \thetable. System Parameters and Model Configuration}} \\
\midrule
\multicolumn{2}{c}{\textbf{System Parameters}} \\
\midrule
Peak Throughput      & 261.12 TFLOPS \\
Memory bandwidth  & 3.35 TB/s \\
D2D bandwidth    & 512 GB/s \\
\midrule
\multicolumn{2}{c}{\textbf{Model-Specific Parameters}} \\
\midrule
TP degree (LLaMA-8B)   & 4 \\
TP degree (LLaMA3-70B) & 8 \\
TP degree (OPT-175B)   & 16 \\
\bottomrule
\end{tabular}
\end{table}

\begin{figure}
    \centering
    \includegraphics[width=0.49\textwidth]{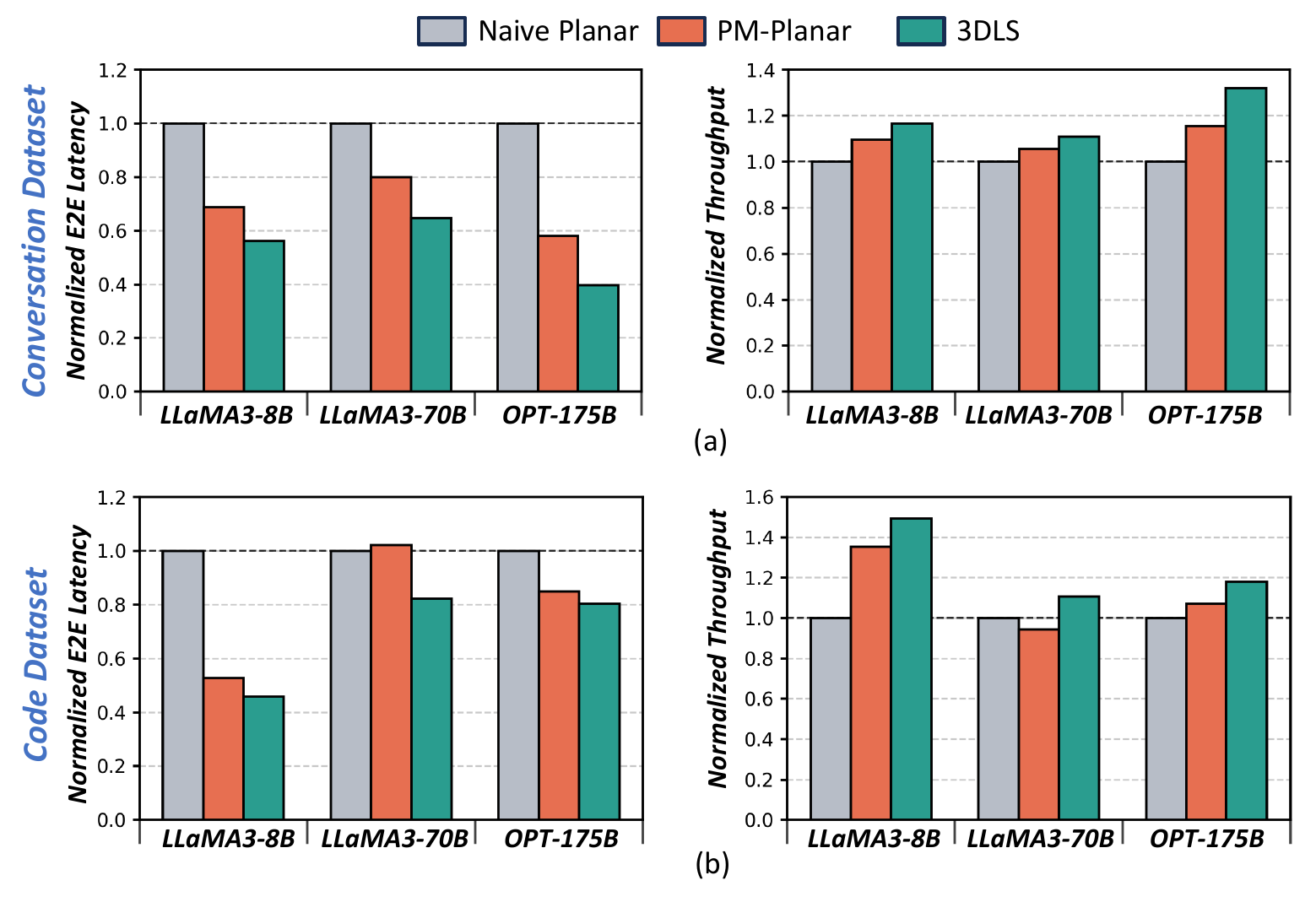}
    \caption{Normalized E2E latency and throughput for Naive Planar, workload-aware PM-Planar, and 3DLS. (a) Conversation Dataset; (b) Code Dataset.}
    \label{fig:fig6}
\end{figure}

\subsection{Evaluation Results and Analysis} 
Fig.~6 shows that PM-Planar improves over Naive Planar in most cases, but 3DLS consistently outperforms PM-Planar. Relative to Naive Planar, 3DLS achieves up to 60.2\% lower E2E latency on Conv/OPT-175B and up to 1.49$\times$ throughput on Code/LLaMA3-8B; across all six model-workload pairs, the geometric-mean gains are 40.6\% lower E2E latency and 1.22$\times$ throughput. Even against workload-aware PM-Planar, 3DLS achieves up to 31.4\% lower latency and 1.17$\times$ throughput, with geometric-mean improvements of 18.2\% lower latency and 1.11$\times$ throughput. The largest latency gain occurs on Conv/OPT-175B at scale=1, PB=16, and DB=64: E2E latency drops from 44.99 s in Naive Planar and 26.08 s in PM-Planar to 17.89 s in 3DLS. The largest throughput gain over Naive Planar occurs on Code/LLaMA3-8B at scale=8, PB=1, and DB=4: throughput rises from 24.32 req/s in Naive Planar and 32.92 req/s in PM-Planar to 36.30 req/s in 3DLS. The largest throughput gain over PM-Planar occurs on Code/LLaMA3-70B at scale=1, PB=1, and DB=1: throughput rises from 5.55 req/s in Naive Planar and 5.24 req/s in PM-Planar to 6.15 req/s in 3DLS.

These results show that priority management is useful but workload-sensitive: the best KVT:AR split differs between Conv and Code, and mismatched static partitions can underperform Naive Planar.  Because PM-Planar still carries KVT and AR on the same physical lateral links, it can only repartition the shared fabric, whereas 3DLS routes KVT vertically and removes it from the lateral collective path.

\section{Related Work}

Recent GPU-based LLM serving systems~\cite{splitwise,distserve} improve serving efficiency through PD disaggregation and phase-aware resource management. \cite{wscllm} extends this direction to wafer-scale serving through co-exploration of serving strategies and wafer-scale architectures. Unlike prior works, we use 3D stacking not merely for bandwidth, but to isolate the KV-transfer path from decode-side collectives in PD-disaggregated LLM serving.

\section{Conclusion}
We identified shared-path contention between layer-wise KV-cache transfer and decode-side TP collectives as a key bottleneck in chiplet-based PD-disaggregated LLM serving.  3DLS removes this interference by separating the KV-transfer path from decode-side TP collevtives. Across the evaluated models and workloads, 3DLS improves serving performance over both Naive Planar and workload-aware PM-Planar, achieving up to 1.49$\times$ throughput and 60.2\% lower E2E latency over Naive Planar, and up to 1.17$\times$ throughput and 31.4\% lower E2E latency over PM-Planar. These results suggest that traffic-class physical isolation is an important architectural principle for future chiplet-based PD-disaggregated LLM serving systems, motivating further co-design of 3D integration, serving orchestration, and hardware organization.

%

\vfill

\end{document}